# A Penetration Depth Study on $Li_2Pd_3B$ and $Li_2Pt_3B$

H. Q. Yuan[1], D. Vandervelde[1], M. B. Salamon[1], P. Badica[2,3], and K. Togano[2]

[1]*Department of Physics, University of Illinois at Urbana & Champaign,
1110 W. Green Street, Urbana, IL 61801, USA*
[2]*Institute for Materials Research, Tohoku University, 2-1-1 Katahira, Aoba-ku, Sendai, 980-8577, Japan*
[3]*National Institute of Materials Physics, Bucharest, POBox MG-7, RO-76900, Romania*

**Abstract.** In this paper we present a penetration depth study on the newly discovered superconductors $Li_2Pd_3B$ and $Li_2Pt_3B$. Surprisingly, the low-temperature penetration depth $\lambda(T)$ demonstrates distinct behavior in these two isostructural compounds. In $Li_2Pd_3B$, $\lambda(T)$ follows an exponential decay and can be nicely fitted by a two-gap BCS superconducting model with a small gap $\Delta_1$=3.2K and a large gap $\Delta_2$=11.5K. However, linear temperature dependence of $\lambda(T)$ is observed in $Li_2Pt_3B$ below $0.3T_c$, giving evidence of line nodes in the energy gap.

## 1. INTRODUCTION

The recent discovery [1,2] of superconductivity in the ternary lithium borides $Li_2Pd_3B$ and $Li_2Pt_3B$ has been attracting wide attention due to their rich physical properties. Most likely, this system provides a model example bridging unconventional superconductivity with the classic BCS superconductivity.

$Li_2Pd_3B$ and $Li_2Pt_3B$ crystallize in a perovskite-like cubic structure composed of distorted octahedral units of $BPd_6$ and $BPt_6$,[3] which resembles other metallic oxides, e.g., the high $T_c$ cuprates and the sodium cobaltates. Measurements of thermodynamic and transport properties revealed a superconducting transition temperature $T_c$ of 7-8K in $Li_2Pd_3B$, but around 2.5 K in $Li_2Pt_3B$.[1,2] The large difference of $T_c$ in these two compounds is difficult to understand within the conventional BCS theory. Up to now, only a very few reports can be found in the literature on the superconducting properties of $Li_2Pd_3B$ and $Li_2Pt_3B$. Similar to $MgCNi_3$, the estimated superconducting parameters of $Li_2Pd_3B$ are close to the BCS values.[1,2] Electronic structure calculations showed that the Fermi surface of these two borides are very complicate and suggested that at least a two-band model is required to study their electronic properties.[4] Recent results of X-ray photoemission spectroscopy demonstrated that the electronic correlation in the Pd-compound does not play a role on the physical properties.[5] Nothing has been reported on the pairing mechanism of these two superconductors and further efforts are required to shed light on the superconducting nature. In this paper, we investigate the penetration depth in these two compounds, which is directly related to their superconducting gap structures.

## 2. EXPERIMENTAL METHODS

Precise measurements of penetration depth $\lambda(T)$ were performed utilizing a tunnel-diode based, self-inductive technique at 21 MHZ down to 90 mK in a dilution fridge. With this method, the change of $\lambda(T)$ is proportional to the resonant frequency shift $\Delta f(T)$, i.e., $\Delta\lambda(T)=G\Delta f(T)$, where the factor G is determined by sample and coil geometries. The magnetization M(T, H) was measured using a commercial SQUID magnetometer (MPMS, Quantum Design).

Polycrystalline $Li_2Pd_3B$ and $Li_2Pt_3B$ have been synthesized by using a two-step arc-melting method. First, alloys of $Pd_3B$ and $Pt_3B$ have been prepared from the mixtures of Pd (99.9%), Pt (99.99%) and B (99.5%). Then Li has been introduced into the melt. Powder X-ray diffraction identifies a single phase. To measure penetration depth, single-crystal-like pieces were selected from the smashed polycrystalline batches.

## 3. RESULTS AND DISCUSSSION

As an example, in Fig.1 the temperature dependence of the frequency shift $\Delta f(T)=f(T)-f(0)$ is shown for $Li_2Pd_3B$. For comparison, the corresponding magnetization M(T) is plotted in the inset. Clearly, the results from different measurements are consistent and demonstrate a sharp superconducting transition around 7K, indicating good sample quality. Similar features are also observed in $Li_2Pt_3B$ with $T_c \approx 2.4K$.

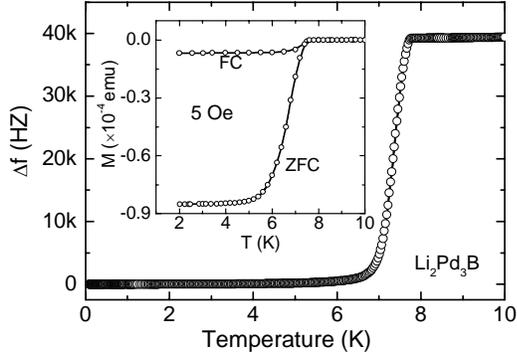

**FIGURE 1.** Temperature dependence of the frequency shift $\Delta f(T)$ for $Li_2Pd_3B$. Inset: magnetization vs. temperature measured in zero field cooling (ZFC) and field cooling (FC).

The low-temperature penetration depth $\Delta\lambda(T)$ of $Li_2Pd_3B$ and $Li_2Pt_3B$, derived from $\Delta\lambda(T)=G\Delta f(T)$ where G=6.28 and 10.13 Å/Hz for Pd- and Pt-compounds respectively, is presented in Fig.2. At first glance, $\Delta\lambda(T)$ shows a much weaker temperature dependence in $Li_2Pd_3B$ than in $Li_2Pt_3B$. A detailed analysis revealed that $\Delta\lambda(T)$ of $Li_2Pd_3B$ can be well fitted by a two-gap BCS model below $0.3T_c$, as described in Ref.[6] for $MgB_2$, with a small energy gap ($\Delta_1$=3.23K) and a large energy gap ($\Delta_2$=11.55K). The fraction $c_1$ contributed from the small gap is about 5%, indicating only a small fraction of carriers are responsible for the opening of the small gap. To explore other possible gap functions, fits to the weak coupling BCS model (dotted line) and to a power law (dash-dotted line) are also included in Fig. 2(a). Obviously, the two-gap model (solid line) is the best fit, suggesting that $Li_2Pd_3B$ is a wholly gapped superconductor. On the other hand, $\Delta\lambda(T)$ of $Li_2Pt_3B$ behaves quite differently at low temperature. Instead of an exponential decrease of $\Delta\lambda(T)$ observed in the Pd-compound, $\Delta\lambda(T)$ follows a linear temperature dependence in the Pt-compound, indicating the existence of line nodes in the superconducting energy gap. It is noted that the superfluid density can be fitted by a two-gap BCS model and a d-wave superconducting gap over the whole temperature range ($T<T_c$) for $Li_2Pd_3B$ and $Li_2Pt_3B$, respectively. Unconventional superconductivity is usually discussed beyond the phonon pairing mechanism. Recently, Agterberge et al argued that exotic superconductivity may arise from the conventional phonon mechanism in the systems with a multi-pocket Fermi surface located at some symmetry points, due to the competition of phonon and Coulomb interactions.[7] Considering the complicate Fermi surface consisting of multiple pockets and the van Hove singularity around the Fermi energy in $Li_2Pd_3B$ and $Li_2Pt_3B$,[4] the evolution of superconductivity and the unconventional superconducting features could be understood within this model; details will be published elsewhere.

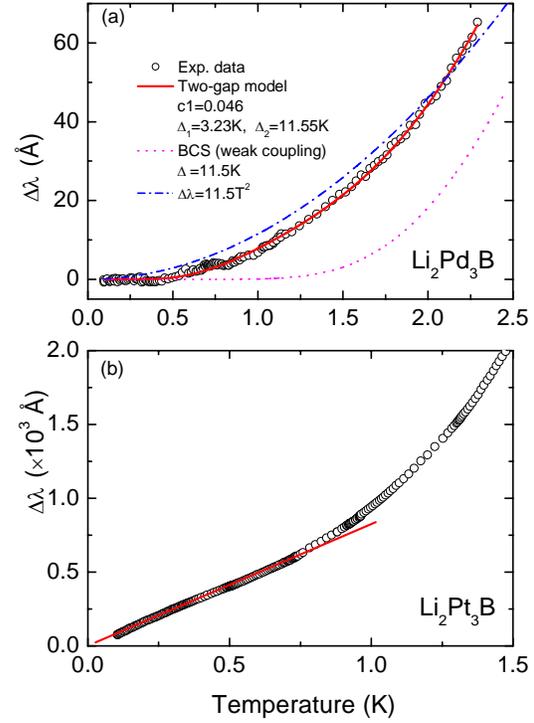

**FIGURE 2.** Temperature dependence of the penetration depth $\Delta\lambda$ (T) for (a) Li2Pd3B and (b) Li2Pt3B.

## ACKNOWLEDGMENTS

We are grateful for the useful discussion with D. F. Agterberg. This work is supported by the Department of Energy. HQY acknowledges the ICAM postdoctoral fellowship.